**Gapless Spin Wave Transport through a Quantum Canted-Antiferromagnet**

Hailong Fu[1], Ke Huang[1], Kenji Watanabe[2], Takashi Taniguchi[3], Jun Zhu[1,4]*

**Affiliations**

[1]Department of Physics, The Pennsylvania State University, University Park, Pennsylvania 16802, USA.

[2]Research Center for Functional Materials, National Institute for Materials Science, 1-1 Namiki, Tsukuba 305-0044, Japan.

[3]International Center for Materials Nanoarchitectonics, National Institute for Materials Science, 1-1 Namiki, Tsukuba 305-0044, Japan.

[4]Center for 2-Dimensional and Layered Materials, The Pennsylvania State University, University Park, Pennsylvania 16802, USA.

*Correspondence to: jzhu@phys.psu.edu (J. Zhu)

In the Landau levels of a two-dimensional electron system or when flat bands are present, e.g. in twisted van der Waals bilayers, strong electron-electron interaction gives rise to quantum Hall ferromagnetism with spontaneously broken symmetries in the spin and isospin sectors. Quantum Hall ferromagnets support a rich variety of low-energy collective excitations that are instrumental to understand the nature of the magnetic ground states and are also potentially useful as carriers of quantum information. Probing such collective excitations, especially their dispersion $\omega(k)$, has been experimentally challenging due to small sample size and measurement constraints. In this work, we demonstrate an all-electrical approach that integrates a Fabry-Pérot cavity with non-equilibrium transport to achieve the excitation, wave vector selection and detection of spin waves in graphene heterostructures. Our experiments reveal gapless, linearly dispersed spin wave excitations in the $E$ = 0 Landau level of bilayer graphene, thus providing direct experimental evidence for a predicted canted antiferromagnetic order. We show that the gapless spin wave mode propagates with a high group velocity of several tens of km/s and maintains phase coherence over a distance of many micrometers. Its dependence on the magnetic field and temperature agree well with the hydrodynamic theory of spin waves. These results lay the foundation for the quest of spin superfluidity in this high-quality material. The resonant cavity technique we developed offers a powerful and timely method to explore the collective excitation of many spin and isospin-ordered many-body ground states in van der Waals heterostructures and open the possibility of engineering magnonic devices.

## I. INTRODUCTION

Spin wave (SW) excitations, also known as magnons, offer fundamental insight into the nature of a magnetically ordered system, similar to phonons of a crystal. In an easy-axis ferromagnet (FM), SW excitations are gapped at zero momentum $k = 0$ by the energy cost of flipping a spin. On the other hand, an easy-plane FM or antiferromagnet (AFM) supports linearly dispersed, gapless SW excitations that correspond to an in-plane precession of the order parameter [1-4]. A magnetic system can also form topological spin textures such as a skyrmion [5,6]. These low-energy collective excitations are potentially useful as information carriers. Magnons in AFM materials are particularly attractive given their ultrafast dynamics and low energy dissipation [4,7-10]. Furthermore, theory predicts that gapless magnons of an easy-plane AFM or canted-antiferromagnet can form a Bose-Einstein condensate and transport spin in a superfluid-like manner without dissipation; this topic has gathered intense interest of the spintronic community lately [7,11-18].

Two-dimensional electron systems (2DESs) placed in a magnetic field constitute an important class of quantum magnets. Here, magnetism develops in the Landau levels of non-magnetic materials because strong electron-electron interaction leads to spontaneously broken spin and isospin symmetries [5,19-25]. This phenomenon is known as quantum Hall ferromagnetism (QHF). QHF gives rise to skyrmions in semiconductor 2DESs [5]. Graphene materials enrich the physics and phenomenology of QHF by introducing isospins such as valley or layer/sublattice [19,20,22]. For instance, the $E = 0$ Landau levels of AB-stacked bilayer graphene is expected to support a spin-valley coherent, canted-antiferromagnetic (CAF) phase with in-plane rotational symmetry [22] and gapless, linearly dispersed SW excitations that correspond to an in-plane precession of the Néel vector [14,16,26-28]. This gapless magnon mode is predicted to support spin superfluidity [12-18]. To date no direct evidence of the CAF order has been obtained. Quantum Hall ferromagnetism and a plethora of spin, valley and charge density wave excitations also manifest in a growing family of magic-angle-twisted van der Waals bilayers where flat bands are formed without a magnetic field [29-32].

Probing the low-energy collective excitations of a QHF system has been experimentally challenging. Scattering techniques used to measure the excitations of bulk magnets, e.g. inelastic neutron or Brillouin light scattering [4,7,33,34], do not function readily at the low temperatures where QHF typically occurs. Specialized techniques, e.g. surface acoustic waves, have been developed to study magnetic excitations of a semiconductor 2DES [35-37]. Nonetheless, accessing the dispersion $\omega(k)$ over a range of $k$ remains difficult and the microscopic size of van der Waals heterostructures presents additional challenge to any spectroscopic technique. New experimental approaches are needed to explore the rich physics, phenomena and technological potential of van der Waals quantum magnets.

In this work, we demonstrate an all-electric method to probe SW excitations in graphene heterostructures, including the attainment of the dispersion relation $\omega(k)$. Key to this approach is the integration of a high-quality Fabry-Pérot (FP) cavity into a multi-terminal transport device, which enables the selective excitation of magnons of discrete wave vectors through resonant transmission. We present unprecedented experimental evidence of gapless, linearly dispersed SW excitations in the $E = 0$ Landau level of bilayer graphene, directly validating the theoretically predicted CAF order. The SW propagates coherently with a high group velocity of several tens of km/s, the magnetic field dependence of which agrees well with a hydrodynamic model. We examine intrinsic and extrinsic sources of dissipation by varying temperature and a number of relevant experimental conditions. These results open the door for the quest of spin superfluidity and the development of more complex magnonic device geometries in this high-quality QHF platform. Our experimental method is compatible with a wide range of sample geometry and measurement conditions. We envision its applicability to other symmetry-broken magnetic ground states in van der Waals materials.

## II. DEVICE FABRICATION AND CHARACTERIZATION

Our bilayer graphene devices are fabricated using van der Waals dry transfer, side contact and precision alignment techniques [38-40] with three layers of gates (5 in total). Figs. 1(a) and (b) show the optical micrograph and schematic sideview of device 606. In areas Q3 and Q4, the filling factor $\nu$ and displacement field $D$ are controlled by the aligned top and bottom gates. The bottom gates are etched into the profile of a quantum point contact. We use this geometry to control areas Q3 and Q4 separately. The bulk filling factor $\nu_B$ in the rest of the device is controlled by the global graphite gate. All contacts reside outside the range of the global gate and are heavily doped by the Si backgate. The use of two long contacts adjacent to the dual-gated region is a salient feature of our design that promotes SW excitation and detection, as we will show below. In a magnetic field, we observe well-developed integer quantum Hall effect (IQHE) in the bulk and the dual-gated regions. Appendix A gives a detailed description of the fabrication steps of devices 606 and 611 and the characteristics of device 606.

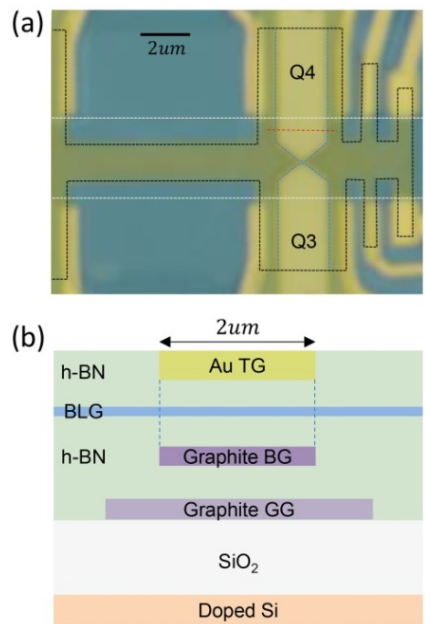

Fig. 1. (a) An optical image of device 606. The black, blue and white dashed lines outline the edges of the bilayer graphene sheet, the graphite bottom gate, and the graphite global gate respectively. Areas Q3 and Q4 are gated by aligned top and bottom gates. The opening of the quantum point contact is 106 nm. Side contacts are made to the bilayer graphene sheet gated by the 295 nm $SiO_2$/doped Si backgate. (b) A schematic side view along the red dashed line in (a).

## III. NON-LOCAL MEASUREMENT SETUP

Figure 2 shows the non-local measurement setup we use to electrically excite and detect spin wave transmission through area Q4. In this setup, Q4 is set to the CAF phase of $\nu = 0$ while Q3 is set to the layer-polarized (LP) phase, which is non-magnetic and insulating [22,24,25,40,41]. Conduction through the quantum point contact is pinched off (See Fig. 8 in appendix A). We note that neither the CAF nor the LP phase carries edge states [22,24,25,40,41] and because of the large width of the insulating middle region ($w$ = 2 μm for device 606 and 1.6 μm for device 611), no edge state on the left could directly transmit to the right [42,43]. The bulk is set to filling factor $\nu_B = 2$, which is fully spin polarized [41]. We apply a varying dc and a small fixed ac voltage $V_{dc} + \delta V_{ac}$ between contacts 2 and 3 and measure a non-local ac voltage signal $\delta V_{NL}$ as a function of $V_{dc}$ on the right side of Q4, e.g. between contacts 8 and 7. We examine the differential non-local signal $dV_{NL}/dV \equiv \delta V_{NL}/\delta V_{ac}$ and integrate $dV_{NL}/dV$ over $V_{dc}$ to obtain $V_{NL}$. Figs. 2(a) and (b) show respectively the flow of the high chemical potential (red) and low chemical potential (blue) edge states under different dc bias conditions. Previous studies examining SW excitations of a spin polarized quantum Hall state have detected appreciable non-local $dV_{NL}/dV$ in similar measurement setups [44,45].

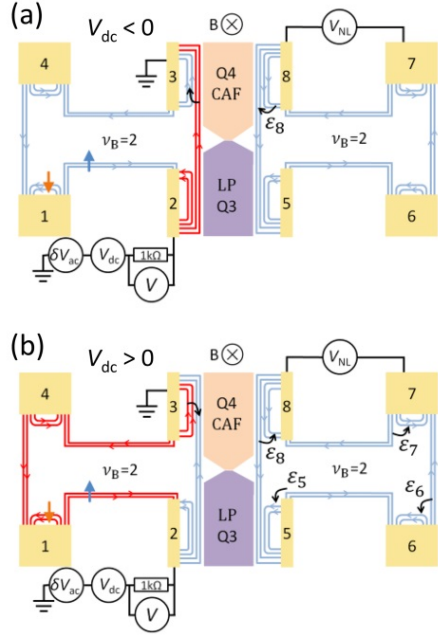

Fig. 2. The non-local differential voltage measurement setup used in Fig. 3(a) and (b) corresponding to negative and positive dc bias conditions respectively. We apply a varying dc bias $V_{dc}$ and a small ac bias ($\delta V_{ac}$ = 10 μV, $f$ = 17 Hz) between two contacts on the left side of the device and measure a non-local ac differential voltage $\delta V_{NL}$ on the right side of the device. A 1 kΩ resistance is used to monitor the ac current. Electrons

riding on edge states departing from a negatively/positively biased contact acquire a high/low chemical potential and are represented in red/blue lines with the chirality given by the magnetic field. Blue and orange arrows indicate the polarization of spin in the bulk and near the contacts, where heavy doping by the Si backgate gives rise to additional edge states. $\varepsilon_n$ indicates chemical potential redistribution at a contact caused by magnon absorption.

## IV. RESULTS AND DISCUSSIONS

We first examine the scenario in which both the bulk and area Q4, called the middle region from now on, are set to $\nu = 2$, where spins are polarized along the external field direction. Figure 3(c) plots the non-local differential signal $dV_{NL}/dV$ and the integrated $V_{NL}$. Both exhibit a $V_{dc}$ threshold of approximately the Zeeman energy $E_z = g\mu_B B$ ($g = 2$). This is because SW excitation in an easy-axis FM is gapped by the spin-flip energy cost of $E_z$. Further, because a fully spin-up polarized FM can only support the propagation of spin-down SWs, the non-local $V_{NL}$ carries the same sign for both positive and negative $V_{dc}$'s. This is indeed what Fig. 3(c) and previous studies at the $\nu = 1$ of monolayer graphene showed [44,45]. While only spin-down SW are emitted, the emission occurs at source (drain) contacts respectively for $V_{dc} < 0$ ($> 0$) [44].

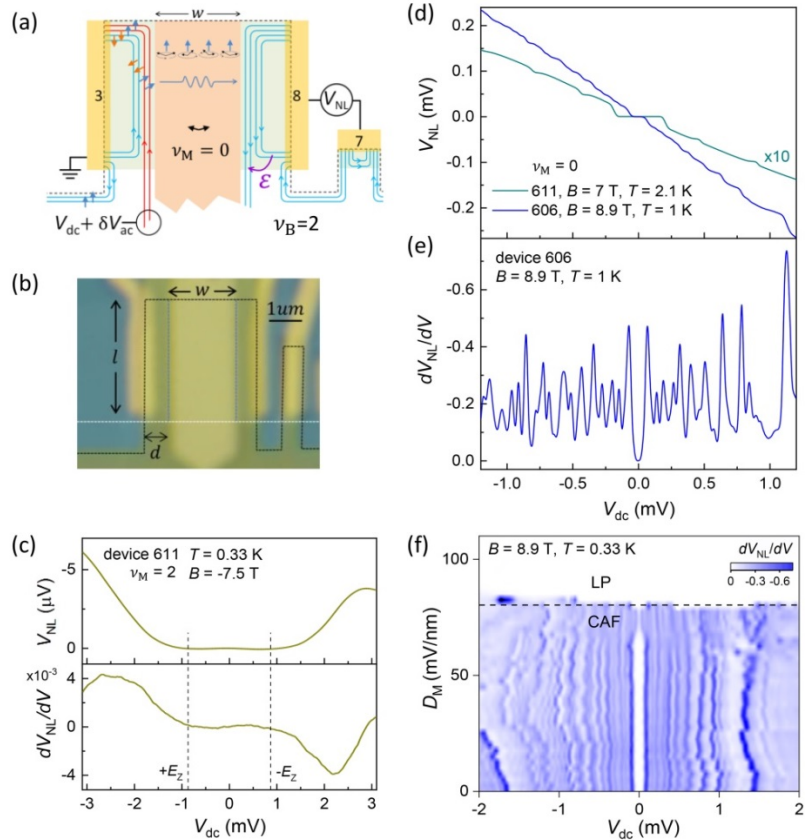

Fig. 3. Excitation, transmission and absorption of gapless SWs through a quantum Hall canted antiferromagnet. (a) Two edge states emitted by a contact below (not shown) and carrying high chemical potential $-eV_{dc}$ scatter with edge states local to the heavily doped contact region shaded in green, emitting SW

with net spin up into the CAF phase. SW transmits through the CAF phase through a precession of the Néel vector. Absorption at a probe contact leads to a chemical potential redistribution $\varepsilon_i$, which is measured in $V_{NL}$. (b) An optical image of the illustrated area in device 606. $l$ = 3.5 μm, $w$ = 2 μm and $d$ = 0.8 μm. (c) $dV_{NL}/dV$ and integrated $V_{NL}$ with $\nu_M = 2$. The gray dashed lines mark $E_z = g\mu_B B$ ($g = 2$) = 0.87 meV. From device 611. (d) $V_{NL}$ ($V_{dc}$) in two devices with the middle region tuned to the CAF phase. Both show an onset $V_T$ much smaller than $E_z$ = 1.0 meV in this set up. Intrinsic and extrinsic contributions to $V_T$ are discussed in Appendix E. (e) The differential signal $dV_{NL}/dV$ in device 606. (f) A false-color graph of $dV_{NL}/dV$ as a function of the $D$-field of the $\nu_M = 0$ region. $D^* \sim 80$ mV/nm separates the CAF phase at low $D$ and the LP, non-magnetic phase at high $D$. The non-local signal is only detected in the CAF phase. Similar behavior is observed in device 611 (Appendix C).

A qualitatively different behavior is expected for an easy-plane AFM or CAF state, which transmits magnons through the in-plane precession of the Néel vector. Both spin-up and spin-down SWs can be transmitted and translate into $V_{NL}$ of opposite signs in the detection region. In a system with in-plane U(1) symmetry, the $k = 0$ mode is expected to be gapless [14,26-28].

Indeed, the characteristics of the non-local signal changes drastically when the middle region of our device is tuned to the putative CAF phase of $\nu = 0$. The results are shown in Figs. 3(d) and (e). Here, $V_{NL}$ varies approximately linearly with $V_{dc}$, changes sign at $V_{dc} = 0$ and commences at $|V_{dc}|$ much smaller than $E_z$. For example, the low-temperature threshold $V_T$ is only 0.025 mV in device 606 while $E_z$ = 1.0 meV. Further, the non-local signal abruptly disappears when $\nu = 0$ transitions to the non-magnetic LP phase at large $D$-field, indicating the necessity of a magnetic order in its detection (Fig. 3(f)). These observations are generally consistent with the transmission of gapless SWs through a CAF phase.

The $dV_{NL}/dV$ signal in our device reaches up to 0.5 with an average of ~ 0.2 (Fig. 3(d) and (e)). It is very large compared to prior graphene devices [44,45] and several orders of magnitude larger than the heavy metal/magnetic insulator interfaces studied in Refs. [8-10]. We attribute the large signal to the unique design of our devices shown in Figs. 3(a) and (b). Heavy doping by the Si backgate leads to the crowding of many edge states in the green shaded area (Fig. 3(a)). This doping profile causes a rapid decrease of the carrier density adjacent to the $\nu_M = 0$ region, which facilitates strong inter-edge scattering. Under a negative dc bias an electron scattered from a "hot" edge carrying canted up-spin to a "cold" edge carrying canted down-spin releases net spin-up angular momentum into the CAF region. Similarly, net spin-down angular momentum is released with a positive dc bias. This process can occur at $E \ll E_z$ due to gradual spin reorientation in the vicinity of the CAF region (Fig. 9 in Appendix B and [14]). Thus, only contacts adjacent to a CAF region can emit gapless SWs (Transmission of gapped SWs through a FM/CAF junction is discussed in Appendix H). In our device, this process occurs along the entire length of contact 3, which is several μm long, in comparison to individual "hot spots" in conventional Hall bar structures [44]. The absorption of SW is dominated by contact 8 on the right side of the CAF region; this causes $dV_{NL}/dV$ to be roughly symmetric in $V_{dc}$, as our data in Fig. 3(e) shows.

A more quantitative description of the above process is given in Appendix B, where we calculate $V_{NL}$ using the spin chemical potential redistribution method introduced in Ref. [44].

Our analysis can satisfactorily explain the sign, symmetry and magnitude of the non-local signal we observed in a variety of measurements using different contact configurations and in magnetic field of both directions.

More strikingly, highly reproducible oscillations in $dV_{NL}/dV$ and corresponding step-like features in $V_{NL}$ develop at low temperatures (Figs. 3(d) and (e)) and also additional data on device 611 in Appendix C). They are strongly reminiscent of discrete standing waves of a confined geometry, e.g. the resonant transmission of a FP cavity [46,47]. Here the FP cavity is that of the SWs. The structure of our devices – a dual-gated region sandwiched between two parallel contacts – motivated us to consider the scenario of a one-dimensional FP cavity. As illustrated in Fig. 4(a), standard FP resonance produces transmissions at wavevectors satisfying $k_n = n\pi/w$ and corresponding magnon energies $E_n = nE_1$, where the fundamental mode $E_1 = \hbar v_{af}\pi/w$. Here $v_{af}$ is the velocity of the SW and $n = 1, 2, 3…$labels the mode number. The $n^{th}$ harmonic manifest as peaks in $dV_{NL}/dV$ at the corresponding dc bias $V_n = \pm E_n/e$. The large amplitude of the oscillations suggests long SW dephasing length of many micrometers in our devices, a necessary condition to explore spin superfluidity [14,16]. We focus on signals within $\pm E_z$ since in this range only excitations on the gapless CAF SW branch are allowed [26,27].

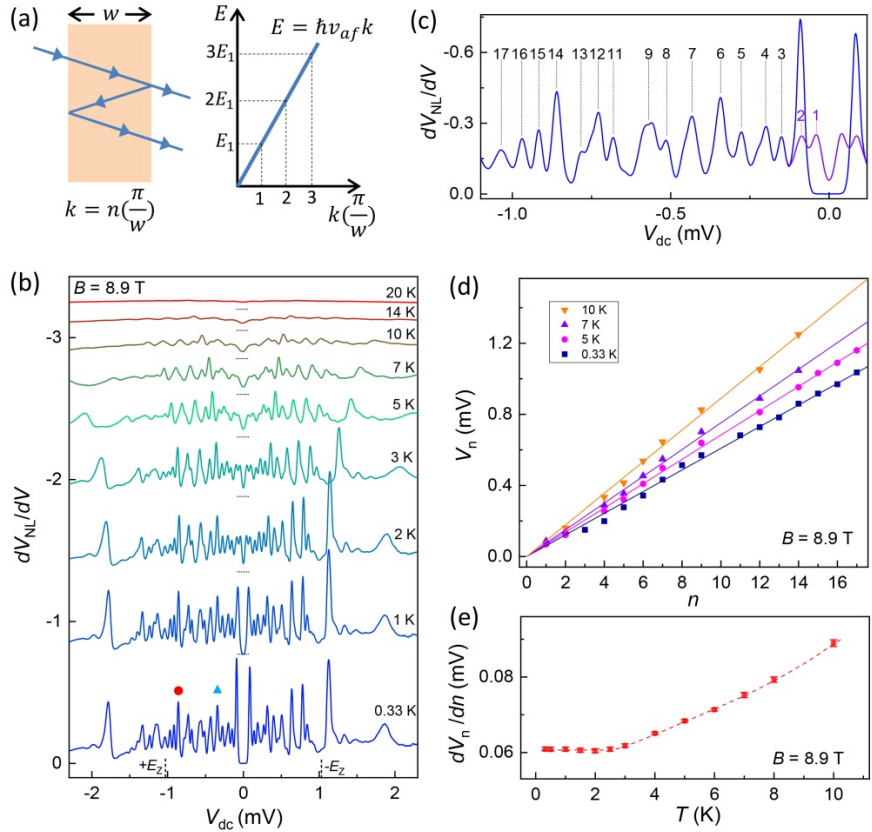

Fig. 4. Fabry-Pérot resonance of a CAF cavity. (a) illustrates the resonant selection of wavevector $k_n$ and corresponding discretization of a linearly dispersed SW. (b) $dV_{NL}/dV$ vs $V_{dc}$ at selected temperatures. Traces are vertically stacked for clarity. Short black dashed lines mark the zero-signal position of each trace. (c) plots $dV_{NL}/dV$ at $T = 0.33$ K (blue) and 2 K (purple) with the mode numbers labeled in the plot. We probe the long

wavelength limit where $k_n l_B < 0.23$ for the entire range. The $n = 1$ and 2 modes are suppressed at low temperatures, likely due to a small contact barrier (Fig. 14(a) of Appendix E). Irregularities of the oscillations are attributed to non-uniformity of the cavity, similar to Ref. [46]. (d) $V_n$ vs mode number $n$ at selected temperatures. Solid lines are fits to data that pass through the origin. Free parameter fitting yields y-axis intercepts less than ½$E_1$ at all temperatures. (e) plots the $T$ dependence of the slope $dV_n/dn$ extracted from the fitting. $B = 8.9$ T in all figures. From 606. Analysis of the FP resonance in device 611 is given in Appendix C.

In Fig. 4(b) we plot $dV_{NL}/dV$ vs $V_{dc}$ at selected temperatures from 0.33 to 20 K with an expanded version of the 0.33 K data shown in Fig. 4(c), together with the labeling of the harmonics $n = 1$ to 17. Figure 4(d) tracks $V_n$ at different temperatures. For each mode, $V_n$ remains a constant at $T < 2$ K and increasingly shifts to larger value at higher temperature. At each temperature, $V_n$ vs $n$ is well described by a linear fit through the origin, which validates the FP resonance model.

The constant slope of $dV_n/dn = 0.06$ mV/mode we obtained at $T < 2$ K (Fig. 4(e)) yields a velocity of $v_{af} = 57$ km/s at $B = 8.9$ T. A $T$-independent SW velocity at low temperatures is in excellent agreement with the hydrodynamic theory of SW [3]. In conventional AFM materials, anisotropy often leads to the opening of a gap at $k = 0$ [4]. Although $k = 0$ is not accessible in our experiment due to the finite size of the CAF region, fitting the data without constraints yields y-axis intercepts less than ½ $E_1$ at all temperatures, from which we estimate an upper bound of $E_0 = 30$ μeV for a possible gap opening at $k = 0$. The small value of $E_0$, which is roughly 3% of $E_z$, indicates that the CAF phase of bilayer graphene has a nearly ideal easy-plane Néel order. This is perhaps not surprising given the vanishing spin orbit coupling and lack of crystal fields in graphene. This character, together with the high quality of graphene devices, makes this system an ideal platform to explore spin superfluidity [14,16].

In the hydrodynamic theory of an easy-plane AFM, the SW velocity $v_{af}$ is given by $v_{af}^2 = \rho_s \chi_z^{-1}$, where $\rho_s$ is the spin stiffness constant that represents the energy cost of in-plane rotational misalignment between neighboring spin sublattices and $\chi_z^{-1}$ is the inverse transverse susceptibility that characterizes the preference of spin lying in the x-y plane [2]. Specifically a Hartree-Fock description produces $v_{af} = 2l_B \sin\theta_s \sqrt{|u_\perp| \tilde{u}}$ for bilayer graphene [26] (Eq. 1), where $l_B = \sqrt{\frac{\hbar}{eB}}$ is the magnetic length, $\theta_s$ is the spin canting angle measured from the z-axis, $\tilde{u}$ is a renormalized energy scale of the CAF phase and $u_{\perp/z}$ is the anisotropy energy in the x-y plane/z direction. The interplay of $u_z$, $u_\perp$, $E_z$, and the valley anisotropy energy $E_v$, which is proportional to the applied $D$-field, gives rise to multiple phases in the $\nu = 0$ Landau level of bilayer graphene [22]. The easy-plane CAF phase occurs at small $E_v$ and $E_z$, together with $u_z > -u_\perp > 0$. Using experimental parameters of Refs. [21,40,41], we obtain $\tilde{u} = 6$ meV, $u_z \approx E_v = 10.4$ meV, $u_\perp \approx -\frac{1}{7} u_z = -1.5$ meV, and a spin canting angle of $\theta_s \approx 70°$ at $B = 8.9$ T (See Appendix F for a detailed analysis). Eq. (1) gives an estimated $v_{af} = 74$ km/s, in excellent agreement with $v_{af} = 57$ km/s obtained in our experiment.

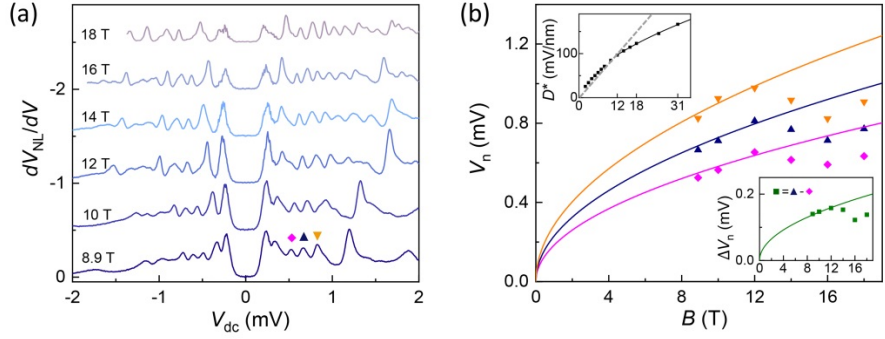

Fig. 5. Magnetic field dependence of the SW signal. (a) plots $dV_{NL}/dV$ at selected magnetic field from 8.9 T to 18 T obtained in the Maglab (See Appendix D for the impact of noise on the measurement done at the MagLab). (b) The main panel plots $V_n$ of three well-reproduced peaks marked in (a). The lower inset plots a differential $\Delta V_n$. Solid lines in both plots show $\sqrt{B}$ scaling. $V_n$ follows the $\sqrt{B}$ scaling below 12 T and tends towards saturation at higher field. The upper inset plots the CAF/LP transition $D^*$ vs $B$ obtained in Ref. [41] with linear (gray dashed line) and $B^{0.56}$ (black solid line) scaling.

The magnetic field dependence of $v_{af}$ further corroborates the above analysis. Here, all interaction energies $\tilde{u}$, $u_z$, $u_\perp$ are proportional to the valley anisotropy energy $E_v^*$ at the CAF/LP phase transition point. $E_v^* = 0.13\ D^*$ is given by the transition field $D^*$ [41]. Experimentally $D^*$ is approximately linear in $B$ at $B < 12$ T, and follows an empirical power law of $B^{0.56}$ above 12 T (upper inset of Fig. 5(b)). The $B$-dependence of $D^*$ leads to a $\sqrt{B}$ dependence of $v_{af}$ at $B < 12$ T and approximately no dependence at higher field. Figure 5(a) plots measurements of $dV_{NL}/dV$ obtained at $B$-fields ranging from 8.9 T to 18 T and Fig. 5(b) tracks the field evolution of $V_n$ and $\Delta V_n$ at several resonances, together with $\sqrt{B}$ trendlines plotted for comparison. Though the data points have considerable spread, they are consistent with the two-segment scaling prescribed by $D^*(B)$. This behavior provides further support to the linear dispersion relation of the SW.

At low temperatures ($T < 2$ K) and when the bulk filling factor is in the range of $1.8 < \nu_B < 2.2$, the non-local signal in our devices reaches a steady state in amplitude and in the resonant $E_n$. These observations suggest that under these conditions the entire process of SW emission, transmission and detection is elastic, i.e. energy conserving, in our devices. Indeed, a constant spin wave velocity $v_{af}$ as $T$ approaches zero is in excellent agreement with theory [3]. Deviations from this steady state occur as $T$ is raised above 2 K or when $\nu_B$ departs from the spin-polarized $\nu_B = 2$ (Fig. 16(a) in Appendix G). Figure 6(a) shows two representative $T$-dependence of the normalized $dV_{NL}/dV$ peak height. Both the $n = 6$ and the $n = 14$ harmonics remain steady at low temperatures, drop precipitously above 2 K and vanish at 10 - 20 K, with the $n = 14$ mode showing a more rapid decay. In Fig. 6(b), we plot the $T$-dependence of the integrated $V_{NL}$. $V_{NL}$ also decreases with increasing temperature but not nearly as rapidly. For example, while $dV_{NL}/dV$ at $n = 6$ drops by a factor of 100 at 20 K, $V_{NL}$ at the same dc bias drops only by a factor of 3.

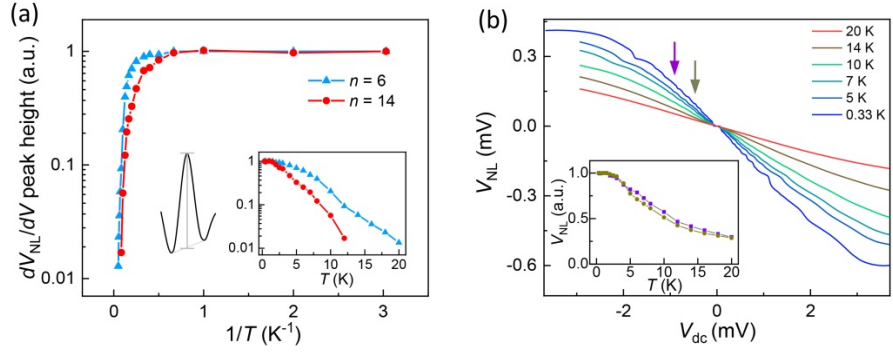

Fig. 6. Temperature dependence of the SW signal. (a) Normalized $dV_{NL}/dV$ peak height as illustrated vs $T$ in an Arrhenius plot for mode $n = 6$ (up-triangle) and 14 (circle) as labeled in Fig. 4(b). Height = 1 at 0.33 K. The inset compares the two modes on a linear $T$ scale. (b) plots the integrated $V_{NL}$ ($V_{dc}$) at selected temperatures. The inset plots normalized $V_{NL}$ ($T$) at two $V_{dc}$'s marked by the arrows.

The strong $T$-dependence our $dV_{NL}/dV$ data in Fig. 6(a) shows can potentially be explained by the occurrence of thermally activated magnon-phonon and magnon-magnon scattering events and their dependence on the incident magnon momentum [2,3,48]. These processes cause decoherence of the SW and thus a rapid decay of the FP resonances and the redistribution of spectral weight to all modes. Thus, their impact on the integrated $V_{NL}$ is much smaller. Increasing temperature may also reduce the transmission of the SW through the now weakened CAF phase. However, this loss is partially compensated by the growing population of thermally excited magnons. These competing factors could account for the much gentler decay of $V_{NL}$ with $T$, as our data in Fig. 6(b) shows.

At elevated temperatures, fits to our data suggest an apparent *increase* of $dV_n/dn$ or $v_{af}$ with increasing $T$ (Fig. 4(e)), which is opposite to trends observed in conventional FM and AFM materials [33,48,49]. Further departure of $\nu_B$ from the spin polarized $\nu_B = 2$ also leads to a systematic blueshift in $V_n$ (Fig. 16(a) in Appendix G), similar to the effect of raising temperature. In the latter measurement, all conditions of the CAF region are held constant so that the blueshift of $V_n$ must originate from external mechanisms. Similar blueshift of $V_n$ is observed when the magnetic field is lowered (Fig. 16(d) in Appendix G). Taken together, these measurements point to the opening of energy dissipation channels in the emission and detection of the SWs *outside* the CAF region. We suspect that the spin polarization of the quantum Hall FM plays an important role. A more in-depth understanding requires microscopic modeling of the voltage-to-spin interconversion processes [28]. In Appendix H, we briefly discuss the transmission of gapped SW excitations through a FM/CAF junction created in device 606. Similar to photons, the creation of more sophisticated magnonic devices can help advance the understanding and technological potential of magons in the arena of quantum information transport [7].

## V. CONCLUSION

In summary, we presented the observation and properties of a linearly dispersed, gapless SW

excitation in a quantum canted-antiferromagnet formed in bilayer graphene. Our results offer direct evidence for this predicted magnetic order and pave the path to the explorations of spin superfluidity in this highly coherent many-body system through microwave radiation or Josephson junction effect. The integration of a resonant Fabry-Pérot cavity and the all-electrical approach enabled the studies of low-energy collective excitations inaccessible to previous experimental techniques. We expect our method to be applicable to a wide range of spin and isospin-ordered quantum magnets emergent in van der Waals materials and heterostructures.

## APPENDIX A: MATERIALS AND LOCAL MEASUREMENTS

### 1. Device fabrication

Devices 606 (Fig.1) and 611 (Fig.12) contain three layers of gating. The global gate and the bottom gates are made of multi-layer graphite flakes. The bottom gates are etched into the shape of a quantum point contact (in device 606) or a strip (in device 611) using standard $O_2$ plasma reactive ion etching (RIE) recipe. The devices are fabricated using the following procedure: 1) Transfer of $h$-BN/graphite global gate to $SiO_2$/Si substrate following Ref. [50]. We have employed this method in our recent studies [39,40,51]. 2) Anneal in Ar/$O_2$ atmosphere at 450°C for 3 hours to remove polymer residue from the transfer. 3) Transfer and pattern the graphite bottom gate using e-beam lithography and RIE. 4) Anneal the stack again using the same annealing procedure. 5) Transfer a $h$-BN/BLG/$h$-BN stack. Here the bilayer graphene sheet is bigger than the global gate. 6) Anneal the stack again. 7) For devices 606, we define the Hall-bar structure of the bilayer graphene using e-beam lithography and RIE ($CHF_3$/$O_2$ plasma). 8) Pattern and deposit Cr/Au side contacts [50]. 9) Pattern and deposit Ti/Au top gates that align with the bottom gates using an alignment procedure we developed previously with a typical precision of 10-15 nm [38,39]. For device 611, in step 7) we pattern and deposit the Ti/Au top gate in the shape of a Hall bar + two handles. 8) Etch the $h$-BN/BLG/$h$-BN stack using the top gate as the mask. 9) Pattern and deposit the Cr/Au side contacts. The top gate overhangs the bottom gate by about 165 nm on each side. This creates another resonant condition that manifests in the non-local signal of device 611 shown in Fig. 12.

### 2. Characteristics of device 606

The operation of device 606 employs 6 gates, which are the Si backgate $V_{Si}$, the graphite global gate $V_{GG}$, the top and bottom gates that control areas Q3 and Q4 respectively $V_{TG3}$, $V_{BG3}$, $V_{TG4}$, and $V_{BG4}$ (see Fig. 8(a)). $V_{Si}$ is biased to a large voltage, e.g. 60 V to dope the contact areas unless otherwise mentioned. The bulk carrier density $n$ and filling factor $v_B$ is controlled by the global gate $V_{GG}$. Sweeping $V_{GG}$ at a fixed $B$-field enables us to determine its gating efficiency and examine the characteristics of the bulk bilayer graphene. As Fig. 7(b) shows, IQHE is well developed in the bulk. By measuring the resistance from electrode 2 to 5, i.e. $R_{2-5}$, as a function of $V_{TG3}$ and $V_{BG3}$, we determine the charge neutrality point (CNP) of area Q3 and the $V_{TG3}$ - $V_{BG3}$ relation, which is shown in Fig. 7(c). Similar measurements are performed on area Q4 and the resulting $V_{TG4}$ - $V_{BG4}$ relation is also given in Fig. 7(c). As expected from the stacking process, $V_{TG3}$ behaves similarly to $V_{TG4}$ and $V_{BG3}$ behaves similarly to $V_{BG4}$. In Fig. 7(d), We measure $R_{XY}$(2-8) as a function of $V_{TG3}$ with Q4 at the LP insulating phase of $\nu = 0$. $1/R_{XY}$(2-8) displays a series of well quantized plateaus as $V_{TG3}$ changes. This allows us to determine the gating efficiency of $V_{TG3}$ and also $V_{BG3}$ through the $V_{TG3}$ - $V_{BG3}$ relation. Similar measurements are performed on Q4 to determine the gating efficiency of $V_{TG4}$ and $V_{BG4}$. Table 1 summarizes the gating efficiencies of all 5 gates. In the majority of our measurements, we position Q3 at the LP phase of $\nu = 0$ by applying a large $D_{Q3} = 400$ mV/nm [41] unless otherwise mentioned.

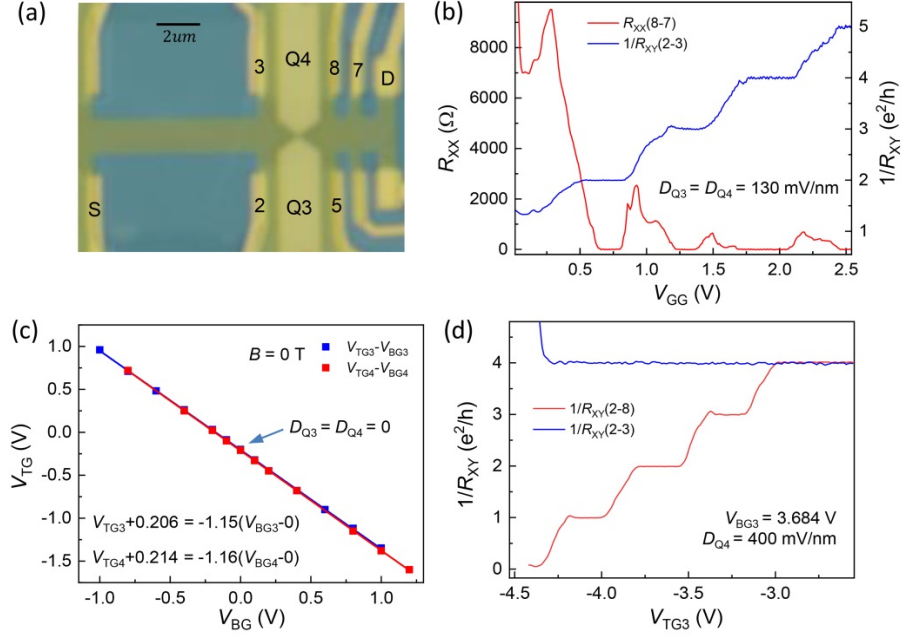

| gate | gating efficiency (cm$^{-2}$ V$^{-1}$) | BN thickness (nm) |
|---|---|---|
| $V_{TG3}$ | 5.15E11 | 32 |
| $V_{TG4}$ | 5.15E11 | |
| $V_{BG3}$ | 5.92E11 | 28 |
| $V_{BG4}$ | 5.98E11 | |
| $V_{GG}$ | 3.45E11 | 48 |

Table 1

Fig. 7. (a) An optical image of device 606. (b) Bulk $R_{XX}$ (to the right of the QPC) and $R_{XY}$ (to the left of the QPC) vs $V_{GG}$ showing well-developed IQHE in the bulk of the bilayer graphene. (c) $V_{TG3}$ – $V_{BG3}$ and $V_{TG4}$ – $V_{BG4}$ relations and the gate voltage offsets corresponding to $D_{Q3} = 0$ (-0.206 V, 0 V) and $D_{Q4} = 0$ (-0.214 V, 0 V). (d) $R_{XY}$ ($V_{TG}$) across area Q3 (red trace). Area Q4 is at the LP insulating phase of $\nu = 0$ with $D_{Q4} = 400$ mV/nm so that $R_{XY}$(2-8) is dominated by Q3. The bulk remains at $\nu_B = 4$ (blue trace). Table 1 summarizes the gating efficiencies of all five gates and the thickness of the h-BN flakes used.

Due to a small misalignment of the top and bottom gates as illustrated in Fig. 8(a), varying $D_{Q3}$ or $D_{Q4}$ while keeping Q3 and Q4 at the CNP also changes the carrier density inside the opening of the QPC. We take advantage of this effect to vary the carrier density inside the QPC opening while holding the bulk filling factor $\nu_B$ constant. The impact of $D_{Q3}$ and $D_{Q4}$ on the QPC is quantified by measuring $R_{2\text{-}8}$ across the QPC at different $D_{Q3}$ or $D_{Q4}$ and tracking the shift in $V_{GG}$ for a point of constant density. Examples are shown in Figs. 8(b) and 8(c) while Fig. 8(d) plots the resulting $V_{GG}$ - $D_{Q3}$ ($D_{Q4}$) relations. The positive sign of the slopes in Fig. 8(d) indicate that both top gates protrude into the QPC opening as shown in Fig. 8(a). For an extended discussion on this subject we refer the reader to Fig. S6 of Ref. [38]. Here we use a large $D_{Q3}$ to pinch off the QPC completely. The operation conditions and the conductance across the QPC are shown in Fig. 8(e). Figure 8(f) illustrates the filling factors and phases in different parts of device 606 under the measurement condition of Fig. 3.

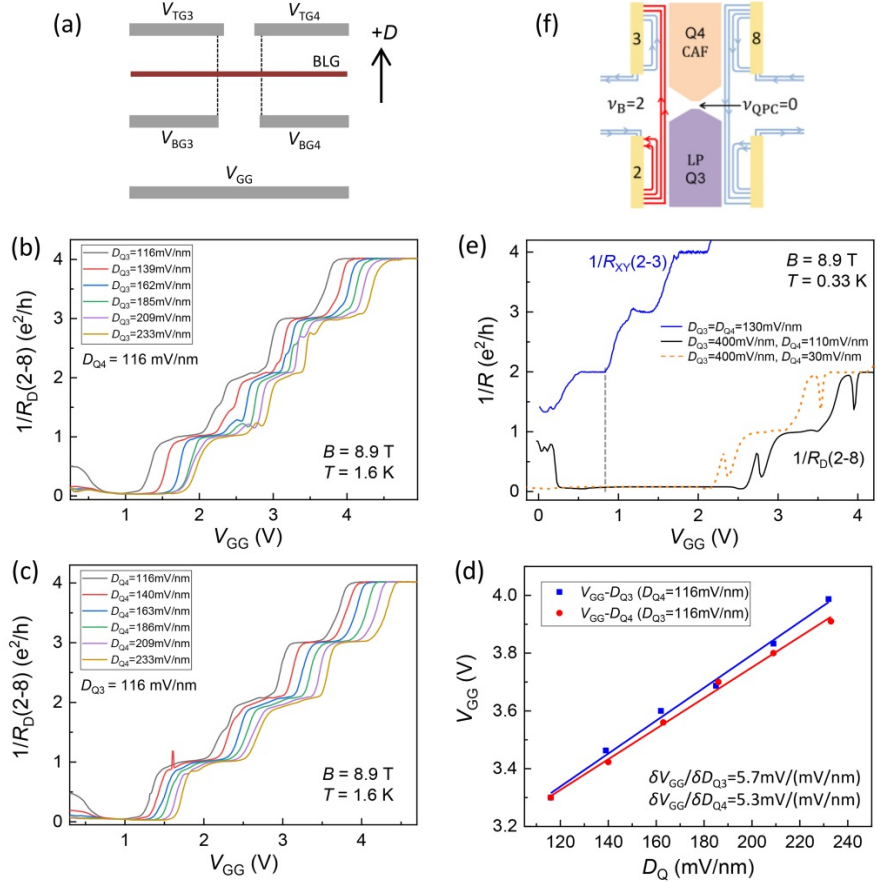

Fig. 8. (a) illustrates the misalignment of the top and bottom gates forming the QPC in device 606. The positive slopes in (d) indicate that both top gates extend into the opening. (b) $1/R_D(2\text{-}8)$ versus $V_{GG}$ at varying $D_{Q3}$. $D_{Q4}$ = 116 mV/nm. (c) $1/R_D(2\text{-}8)$ versus $V_{GG}$ at varying $D_{Q4}$. $D_{Q3}$ = 116 mV/nm. Tracking the shift of the $\nu$ = 3 plateau allows us to obtain the relation between $V_{GG}$ and either $D_{Q3}$ or $D_{Q4}$. The results are plotted in (d). (e) $1/R_D(2\text{-}8)$ versus $V_{GG}$ at $D_{Q4}$ =110 mV/nm and $D_{Q3}$ = 400 mV/nm (black trace). Also plotted is $1/R_{XY}$ in the bulk (blue trace) for reference. In the majority of our non-local measurements, we set $D_{Q3}$ = 400 mV/nm, $D_{Q4}$ = 30 mV/nm and $V_{GG}$ = 0.8 V. In this configuration the bulk is at $\nu_B = 2$, Q3 is in the LP phase, and Q4 is in the CAF phase. We obtain the expected QPC conductance (orange dashed trace) by shifting the measurement at $D_{Q4}$ = 110 mV/nm by -0.42 V according to the gating relation in (d). It shows that the QPC opening sits squarely on the $\nu = 0$ plateau. (f) illustrates the filling factor of different areas and edge states flow in this configuration.

## APPENDIX B: THE NON-LOCAL MEASUREMENT METHOD AND MECHANISMS OF SPIN WAVE EMISSION AND DETECTION

We use Yokogawa GS200 to apply a dc voltage and add a small ac excitation $\delta V_{ac}$ generated by a lock-in amplifier through a transformer and a divider. $\delta V_{ac}$ = 10 µV at 17 Hz is used unless otherwise mentioned. The non-local voltage $\delta V_{NL}$ is measured by a Stanford SR860 lock-in with a NF LI-75A preamplifier. We calculate and present the dc voltage dropped on the sample using $V_{dc} = \frac{R_B}{R_B + R_C} V'_{dc}$, where $V'_{dc}$ is the applied dc voltage, $R_B$ is the sample resistance excluding contacts, and $R_C$ is the total non-sample resistance including the two contacts, cryostat wiring and the 1 kΩ resistor shown in Fig. 2. At a bulk filling factor of 2, $R_B \approx 13$ kΩ and $R_C \approx 1.8$ kΩ in device 606. Both remain constant in our measurements. $V_{dc} = 0.88 V'_{dc}$. In device 611, $R_C \approx 13$ kΩ and $V_{dc} = 0.49 V'_{dc}$ at $\nu = 2$.

In the following we provide qualitative and quantitative accounts of how gapless SW excitations are generated and detected in our non-local measurement. Our descriptions largely follow the model developed by Wei et al for gapped SW excitations [44] and are adpated to the unique characteristics of the CAF phase as illustrated below. Figures 2(a) and 2(b) of the main text compare the non-equilibrium edge state flow in the cases of a negative/positive dc bias. The bulk ($\nu_B = 2$) supports two edge states with spin-up polarization. They depart from contact 2 with high chemical potential $\mu = -eV_{dc}$ with a negative dc bias and terminate at contact 3. In the area between contact 3 and the CAF region, these two edge states with canted up-spins interact strongly with edge states departing and terminating at contact 3 but carrying opposite, canted down-spins as illustrated in Fig. 9. Strong inter-edge transitions launch SWs into the CAF phase from contact 3. A negative $V_{dc}$ launches SW of net spin up, which gives rise to a positive $V_{NL}$. A positive $V_{dc}$ launches SW of net spin down, resulting in a negative $V_{NL}$. Different from a spin polarized quantum Hall state [44], the CAF phase supports the transmission of both types of SWs by further canting of its spin vectors [13].

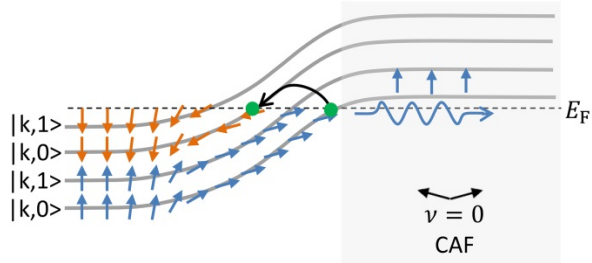

Fig. 9. A schematic of the gradual canting of spins due to the effective field created by the CAF state following Ref. [14]. The canting enables a gapless transition of an electron from the inner edges (quasi spin-up) to the outer edges (quasi spin-down), with the net spin angular momentum carried away by the emission of a SW into the CAF phase. Transition in the opposite direction reverses the net spin angular momentum transmitted and hence the sign of the detected non-local signal.

On the other side of the CAF region, SWs are absorbed by individual contacts, resulting in a chemical potential redistribution $\varepsilon_i$ at each contact. A SW of net spin up cannot propagate in the bulk of $\nu = 2$, which is already fully polarized in the up direction. Thus, all magnons are absorbed by contact 8 next to the CAF region (Fig. 2(a)). While multiple contacts can absorb magnons with net spin down (Fig. 2(b)). Following Ref. [44], we write down the following expressions for the chemical potential at the probe contacts:

For $V_{dc} < 0$:

$$C8: \ 4\mu_8 = 2\mu_7 + 2\mu_8 - \varepsilon_8 \quad (S1)$$
$$C7: \ 4\mu_7 = 2\mu_6 + 2\mu_7 \quad (S2)$$
$$C6: \ 4\mu_6 = 2\mu_6 + 2\mu_5 \quad (S3)$$
$$C5: \ 4\mu_5 = 2\mu_8 + 2\mu_5 + \varepsilon_8 \quad (S4)$$

Therefore,

$$\mu_8 = \mu_7 - \frac{\varepsilon_8}{2} \quad (S5)$$
$$\mu_7 = \mu_6 = \mu_5 \quad (S6)$$

The non-local voltage $V_{NL}(8-7)$ is:

$$V_{NL}(8-7) = \frac{\mu_8}{-e} - \frac{\mu_7}{-e} = \frac{\varepsilon_8}{2e} \quad (S7)$$

The differential voltage $\frac{dV_{NL}(8-7)}{dV}$ is:

$$\frac{dV_{NL}(8-7)}{dV} = \frac{d(\frac{\varepsilon_8}{2e})}{d(\frac{\mu_{sd}}{-e})} = -\frac{1}{2}\frac{d\varepsilon_8}{d\mu_{sd}} \quad (S8)$$

$\mu_{sd}$ is the source-drain chemical potential, which is positive for $V_{dc} < 0$. Equations S7 and S8 produce positive $V_{NL}(8-7)$ and negative $\frac{dV_{NL}(8-7)}{dV}$ for $V_{dc} < 0$, in agreement with data shown in Figs. 3(d) and 3(e).

For $V_{dc} > 0$:

$$C8: \ 4\mu_8 = 2\mu_7 + 2\mu_8 + \varepsilon_8 - \varepsilon_7 \quad (S9)$$
$$C7: \ 4\mu_7 = 2\mu_6 + 2\mu_7 - \varepsilon_6 + \varepsilon_7 \quad (S10)$$
$$C6: \ 4\mu_6 = 2\mu_6 + 2\mu_5 + \varepsilon_6 - \varepsilon_5 \quad (S11)$$
$$C5: \ 4\mu_5 = 2\mu_8 + 2\mu_5 - \varepsilon_8 + \varepsilon_5 \quad (S12)$$

Therefore,

$$\mu_8 = \mu_7 + \frac{1}{2}(\varepsilon_8 - \varepsilon_7) \quad (S13)$$
$$\mu_7 = \mu_6 + \frac{1}{2}(\varepsilon_7 - \varepsilon_6) \quad (S14)$$
$$\mu_6 = \mu_5 + \frac{1}{2}(\varepsilon_6 - \varepsilon_5) \quad (S15)$$
$$\mu_5 = \mu_8 + \frac{1}{2}(\varepsilon_5 - \varepsilon_8) \quad (S16)$$

The non-local voltage $V_{NL}(8-7)$ is:

$$V_{NL}(8-7) = \frac{\mu_8}{-e} - \frac{\mu_7}{-e} = -\frac{1}{2e}(\varepsilon_8 - \varepsilon_7) \quad (S17)$$

The differential voltage $\frac{dV_{NL}(8-7)}{dV}$ is:

$$\frac{dV_{NL}(8-7)}{dV} = \frac{d\left[-\frac{1}{2e}(\varepsilon_8 - \varepsilon_7)\right]}{d(\frac{\mu_{sd}}{-e})} = \frac{1}{2}\frac{d(\varepsilon_8 - \varepsilon_7)}{d\mu_{sd}} \quad (S18)$$

Equations S17 and S18 again produce the correct sign for $V_{NL}$ and $dV_{NL}/dV$ given that $\mu_{sd}$ is negative for $V_{dc} > 0$. Because contact 8 is much larger and much closer to the CAF phase than contact 7, it plays a dominant role in the detection of $V_{NL}$. This makes $V_{NL}$ an approximately odd function of $V_{dc}$ and $dV_{NL}/dV$ an approximately even function of $V_{dc}$, as our data in Figs. 3(d) and 3(e) show.

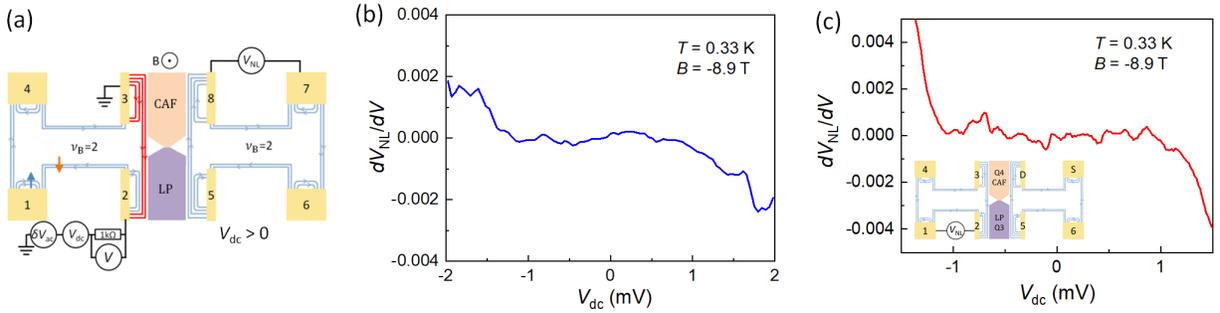

Fig. 10. (a) The same measurement setup as in Fig. 2(b), but with the *B*-field pointing in the opposite direction. All edge states near the CAF phase departure from contact 3 and have the same chemical potential. Electron scattering cannot happen. (b) The measured non-local differential signal using the configuration in (a) is negligible within ± $E_Z$ = 1 meV, in comparison to signal amplitude > 0.1 in the other *B*-field direction. (c) The measured non-local differential signal using the configuration shown in the inset is very small because both contacts 1 and 2 are away from the CAF region.

We have experimented with measurement setups different from that used in Fig. 2. Both the presence of the CAF phase and an active contact in its immediate vicinity (Contact 3 or 8 in Fig. 10(a)) are essential to the emission and detection of the gapless SWs, independent of whether the SW is emitted from the left side or the right side of the sample. In Fig. 10(a), because of the direction of the magnetic field, all edge states near contact 3 are on the same chemical potential so no gapless SW is emitted there and we observe negligible non-local signal in Fig. 10(b). In Fig. 10(c), the voltage probes are some distance away from the CAF region. Because the spin polarized bulk does not support the transmission of gapless SWs, the measured $V_{NL}$ is also close to zero in the range of $|V_{dc}| < E_Z$.

When the middle region is set to $\nu = 2$ (Fig. 3(c)), or when both source and drain contacts are away from the CAF region (Fig. 17), we observe SWs with an excitation gap, similar to what has been reported in the literature. In these scenarios, only SW with net spin down is transmitted and it is emitted by the drain/source contact in the case of a positive/negative dc bias. The non-local $V_{NL}$ (8-7) remains negative and the differential signal $dV_{NL}/dV$ changes sign at zero bias. This is indeed what we observed. A more detailed discussion of gapped SW excitation and transmission is given in Appendix H.

## APPENDIX C: FABRY-PÉROT RESONANCES IN DEVICES 606 AND 611

In Figure 11, we present data and analysis of the Fabry-Pérot oscillations in the $V_{dc} > 0$ regime of device 606. Data from both bias directions overlap strongly and yield very similar results on $v_{af}$. A small difference could be due to a slight bias dependence of the contact resistance or impurity states and geometrical imperfections that respond to positive/negative dc biases differently. An intrinsic asymmetry caused by the detection of the SW can also contribute. See discussions in Appendix B.

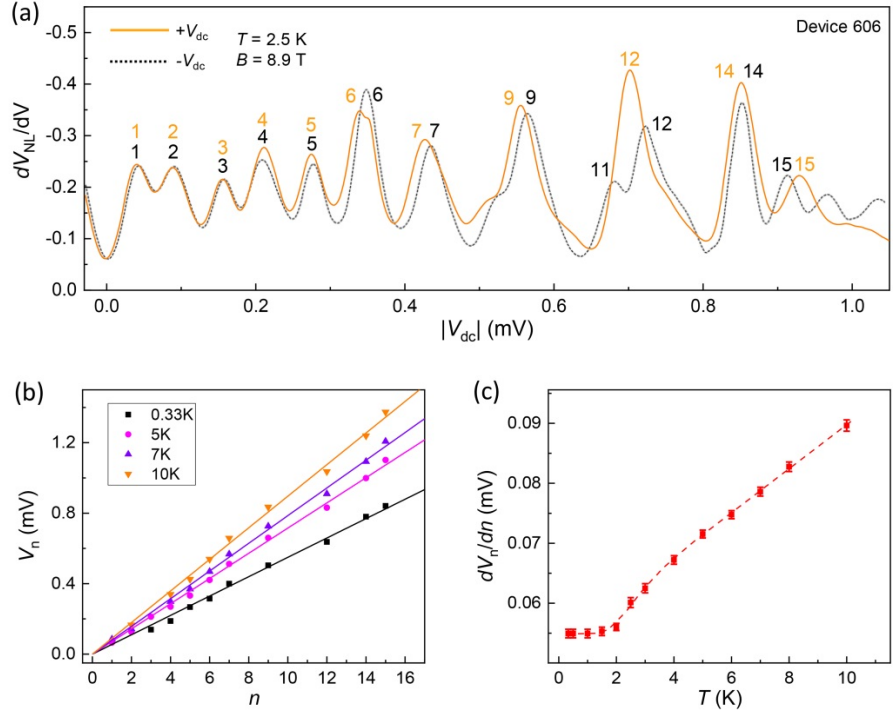

Fig. 11. $dV_{NL}/dV$ and analysis for positive dc bias in device 606. (a) compares $dV_{NL}/dV$ vs positive $V_{dc}$ (orange line) and negative $V_{dc}$ (black dotted line) at $T = 2.5$ K with the resonance peaks labeled for both traces. The two data overlap very well. (b) $V_n$ vs mode number $n$ at selected temperatures and linear fits to the data. (c) Temperature dependence of the slope $dV_n/dn$ extracted from the fittings in (b). The low-$T$ slope of 0.055 mV yields $v_{af} = 52$ km/s, in comparison to $v_{af} = 57$ km/s extracted from the negative $V_{dc}$ data.

Figure 12 presents data from device 611. In this device, the active region supporting gapless SW transport is the area inside the red box in Fig. 12(a), where the dual gated region is positioned at the CAF phase of $\nu = 0$ with $D = -30$ mV/nm. Figure 12(b) presents the properties of the bulk and the measured non-local signal and analysis are shown in Figs. 12(c)-(e). As Fig. 12(a) shows, the top gate is larger than the bottom gate by 165 nm on each side (This was done purposefully). We suspect that this arrangement creates two resonant cavities of width 1.27 $\mu$m and 165 nm respectively. Our data in Fig. 12(d) indeed shows two sets of resonance peaks with very different spacings. The spacing between the sharp resonance peaks is roughly 0.4 mV while the oscillations similar to that in device 606 have a period of 0.059 mV (Fig. 12(c)). The ratio between the two periods, 0.4 mV/0.059 mV agrees well with the inverse ratio of the two cavity widths 1270 nm/165 nm. The linear fitting of the finely spaced oscillations in Fig. 12(e) yields a SW velocity of 36 km/s at 7 T. This value is generally consistent with that of 50-60 km/s obtained in device 606 at 8.9 T given the field difference. Similar to device 606, Fig. 12(f) shows that the non-local signal in device 611abruptly disappears when the dual-gated area transitions to the non-magnetic insulating phase at large $D$-field, again demonstrating the unambiguous role of the CAF phase in the detection of the SW signal.

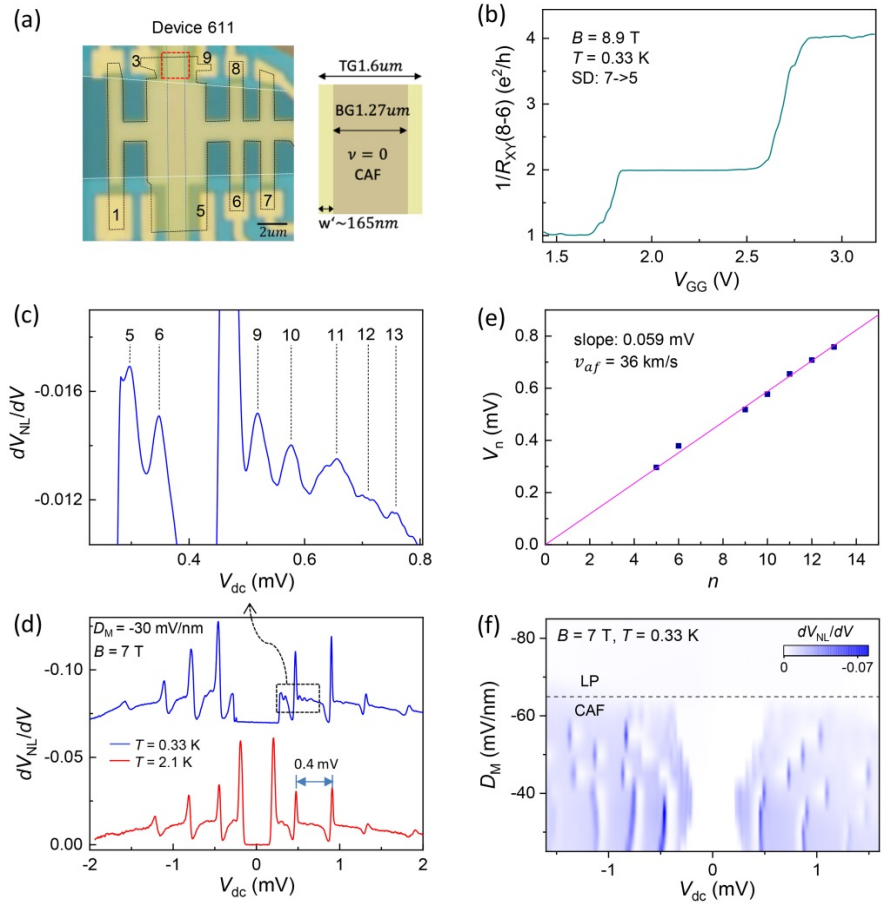

Fig. 12. Characteristics and non-local measurements on device 611. (a) An optical image and a schematic of the active region. The black, blue and white dashed lines outline the edges of the bilayer graphene sheet, the graphite bottom gate, and the graphite global gate respectively. The bottom gate is a strip of width 1.27 μm. The Au top gate includes a "belly" that coincides with the edge of the bilayer graphene sheet and two "handles" that overhang the bottom gate by about 165 nm on each side. (b) $1/R_{XY}(8-6)$ vs the global gate voltage $V_{GG}$ showing well developed IQHE at bulk filling factor $\nu_B = 2$. (c) and (d): $dV_{NL}/dV$ vs $V_{dc}$ at 7 T and two different temperatures showing the two sets of Fabry-Pérot resonances. Contacts 1 and 3 are used as source and drain and the non-local differential voltage is measured from 9 to 8. (e) A linear fit of the fine oscillations similar to that of device 606 yields a slope of 0.059 mV, which corresponds to a SW velocity of 36 km/s. (f)

shows the evolution of the non-local signal as a function of $D_M$ of the dual-gated area in (a). The transition at 65 mV/nm corresponds to the CAF/LP phase transition of $\nu = 0$ [41].

**APPENDIX D: THE IMPACT OF RF NOISE ON THE FABRY-PÉROT RESONANCES**

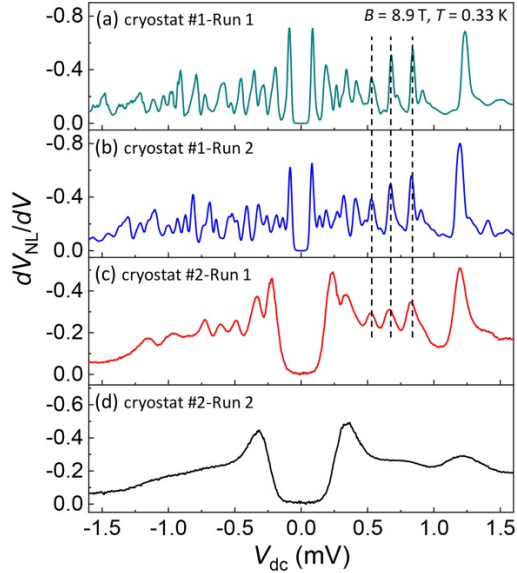

Fig. 13. Sensitivity of Fabry-Pérot resonance to RF noise. (a) and (b) are from a He3 cryostat equipped with RF filtering using thermocoax cables (THERMOCOAX, Inc. Model: 1 Nc Ac 05, length > 2 m). The sample is thermal cycled to above 20 K in between (a) and (b). Resonant peaks in $dV_{NL}/dV$ reproduce very well, apart from occasional small shift. (c) and (d) are obtained at the MagLab using the dilution fridge of the 18 T system (SCM1). They reproduce the envelope of the non-local signal but exhibit larger onset dc bias and substantially fewer resonance peaks. Environmental noise that leads to decoherence is likely the cause. The severity of the problem also depends on what other instruments are running at the same time. The situation in (d) is worse than in (c). From device 606. $T = 0.33$ K. $B = 8.9$ T. Data presented in Fig. 5(a) were obtained in the run of panel (c), where we tracked the magnetic field dependence of the three strong resonance peaks that reproduce well to 18 T.

**APPENDIX E: CONTRIBUTIONS TO THE DC BIAS THRESHOLD $V_T$ IN GAPLESS SW EXCITATIONS**

Although our measurements probe the gapless SW dispersion of the CAF phase, several intrinsic and extrinsic factors give rise to a finite onset bias $V_T$. The first is the finite size of the CAF region, which produces discrete resonant modes at $k_n = n\pi/w$, where $n = 1, 2, 3…$. In addition, the properties of the contact area play an important role in the emission and detection of gapless SWs. Figure 14 gives two examples. Fig. 14(a) shows the temperature dependence of the first three Fabry-Pérot modes in device 606. The first two modes $n = 1$, and 2 are suppressed at low temperatures and only appear at $T > 1.5$ K, probably due to thermal activation over a small contact barrier. Fig. 14(b) plots traces taken on device 611 at four different $V_{Si}$ voltages.

Measurements taken at $V_{Si}$ = 45 and 60 V show the characteristics of gapless SW excitations with a small and stable threshold $V_T$. At lower $V_{Si}$, the data increasingly take on the symmetry and threshold of gapped SWs that propagate through a spin polarized bulk. See Appendix H for an expanded discussion on a FM/CAF junction. These measurements show that heavy doping of the emission/detection contacts, which leads to crowed edge states in their vicinity, is necessary to probe gapless SW excitations.

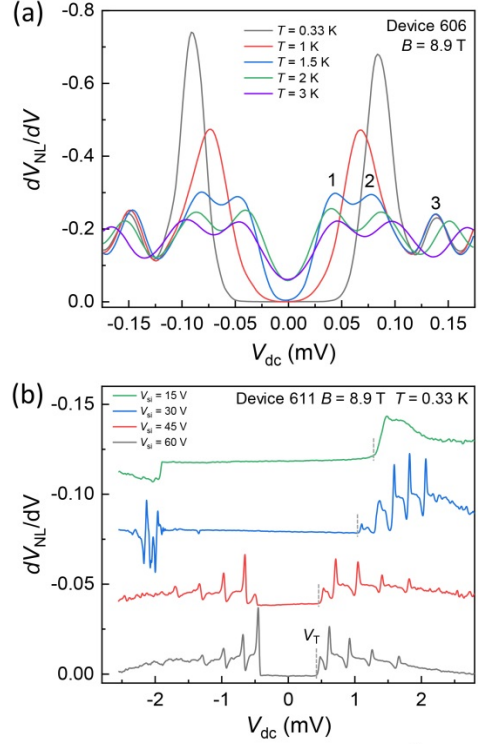

Fig. 14. The threshold bias $V_T$ in the measurement of gapless SW signal. (a) $dV_{NL}/dV$ at different temperatures illustrating the decrease of $V_T$ with increasing temperature. Modes 1 and 2 only appear at $T$ = 1.5 K and above. Modes of $n$ = 3 and above are not impacted by the threshold shift. From device 606. (b) The decrease of $V_T$ with increasing contact area doping. From device 611. See Fig. 12(a) for measurement setup. Traces are vertically shifted for clarity. In both devices, the onset of the non-local signal occurs at $V_T \ll E_Z$ with $V_{Si}$ = 60 V.

## APPENDIX F: THE PHASE DIAGRAM AND ENERGY SCALES OF THE $\nu$ = 0 STATE IN BILAYER GRAPHENE

Previous experiments have shown that all relevant interaction energies and the Landau level gaps of the $E$ = 0 octet in bilayer graphene ($\nu = 0, \pm 1, \pm 2, \pm 3$) scale linearly with $B_\perp$ up to 12 T [21,41]. The $\nu = 0$ phase diagram is driven by competing interaction energies $u_z$, $u_\perp$, $E_z$, and $E_v$. $u_z$ and $u_\perp$ are anisotropy energies in the z-axis and x-y plane respectively. $E_z$ is the Zeeman energy proportional to $B_{tot}$ and $E_v$ is the valley anisotropy energy proportional to the applied $D$-field [22,24,25,40,41]. Small $E_v$ and $E_z$, together with $u_z > -u_\perp > 0$, give rise to the easy-plane

CAF phase studied here. Increasing $E_z$ through $B_{tot}$ drives a transition to an easy-plane FM phase at $u_\perp = \frac{-E_z}{2}$ while increasing $E_v$ drives a transition to a partially layer polarized (PLP) non-magnetic phase at $E_v = \sqrt{u_z^2 - u_\perp^2}$ in a pure $B_\perp$ field [22]. Figure 15 plots a phase diagram from our previous work [40]. At $B_\perp = 3$ T, the CAF-PLP phase transition occurs at $D^* = 34$ mV/nm and the CAF-FM transition occurs at $B_{tot} = 11.2$ T. Together with $E_v$ (meV) $= 0.13\, D\, (\frac{mV}{nm})$ determined in Ref. [41], we obtain $u_\perp = -0.65$ meV and $u_z = 4.4$ meV or $u_\perp \approx -\frac{1}{7} u_z$. The approximately linear $D^*(B)$ relation at low field (inset of Fig. 5(b)) yields $u_z$ (meV) ~ 1.1 $B_\perp$ (T). Measurements of the CAF-FM transition [24,40] also suggest a linear $B_\perp$ scaling for $u_\perp$. This allows us to estimate the corresponding energies at 8.9 T to be: $u_z \approx E_v^* = 10.4$ meV, $u_\perp \approx -\frac{1}{7} u_z = -1.5$ meV. Here we used the measured $D^* = 80$ mV/nm. The spin canting angle $\theta_s$ is given by $\cos\theta_s = \frac{E_z}{2|u_\perp|}$, which is about $70°$ and verifies that indeed the spins lie nearly in the x-y plane. The velocity of the SW in the CAF phase is given by $v_{af} = 2l_B \sin\theta_s \sqrt{|u_\perp|\tilde{u}}$, where $\tilde{u}$ is a renormalized interaction energy at $\nu = 0$ [26]. Following Ref. [26], we use the measured gap of the CAF phase $\Delta_0$ in suspended bilayer graphene [21] to estimate $\tilde{u}$. $\Delta_0$ (meV) $\approx 1.7\, B_\perp$ (T) in Ref. [21] and $\tilde{u} = \frac{89}{224}\Delta_0 = 0.67 B_\perp$ [26]. This gives $\tilde{u} = 6$ meV at 8.9 T. This energy is consistent with the $T$-dependence of the resistance of the CAF phase shown in experiment [40]. It is also reasonable compared to the scaling of $u_z$ and other exchange-dominated energy scales of the system. For example, the $\nu = 2$ gap $\Delta_2$ (meV) $\approx 1.2 B_\perp$ (T) [41,52].

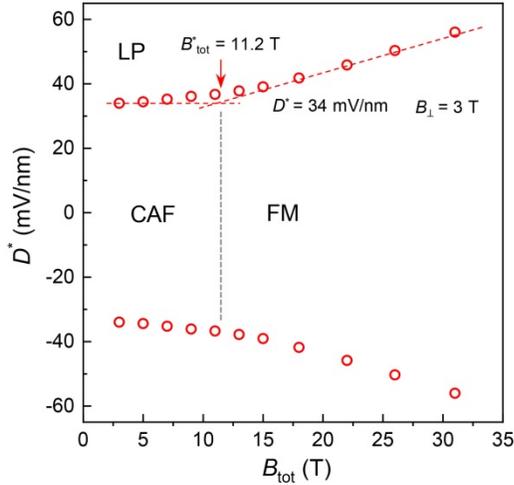

Fig. 15. The phase diagram of $\nu = 0$ in bilayer graphene at $B_\perp = 3$ T showing the CAF-LP transition at 34 mV/nm and the CAF-FM crossover at $B_{tot} = 11.2$ T. Adapted from Ref. [40].

# APPENDIX G: THE BULK FILLING FACTOR AND MAGNETIC FIELD DEPENDENCE OF THE NON-LOCAL $dV_{NL}/dV$

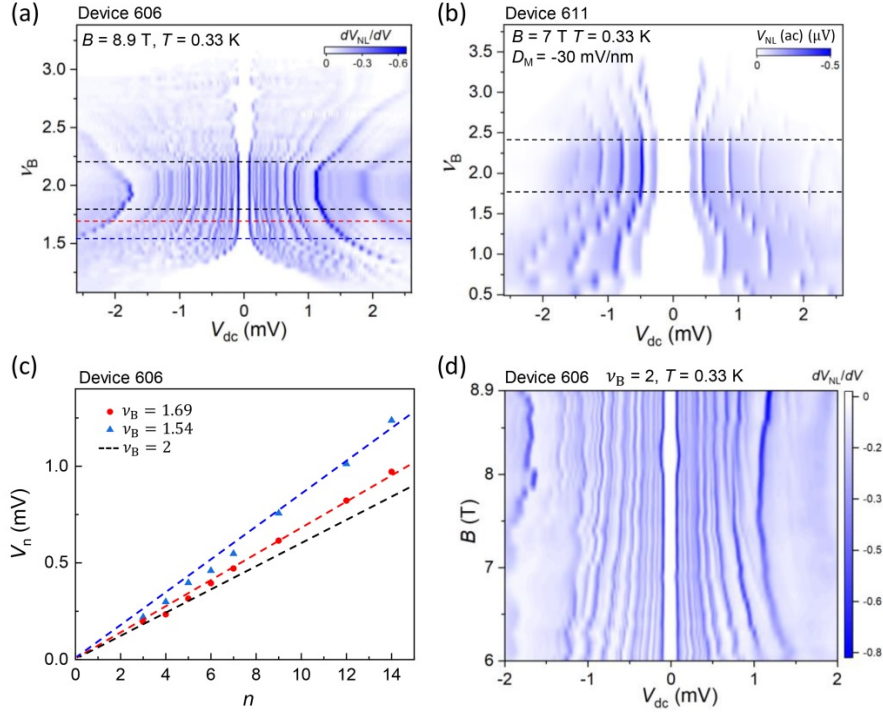

Fig. 16. The dependence of $dV_{NL}/dV$ on the bulk filling factor $\nu_B$ and the external magnetic field with a constant gating condition of the CAF phase. (a) and (b) show the evolution of $dV_{NL}/dV$ as a function of $\nu_B$ in device 606 and 611 respectively. The non-local signal is insensitive to the change of $\nu_B$ in the filling factor range of $\nu_B = (2 \pm 0.2\text{-}0.3)$ (regions within the black dashed lines in both graphs). Further deviation of $\nu_B$ leads to amplitude reduction and simultaneous blueshift of resonant mode energies. The details are sample dependent. (c) characterizes the shift of the mode energy at two different filling factors as indicated by the blue and red dashed lines in (a). Larger shifts are observed for higher harmonics. $\Delta V_n$ is roughly linear in $n$, as the dashed lines show. (d) shows the evolution of $dV_{NL}/dV$ as a function of $B$ while keeping $\nu_B = 2$. Blueshift of resonant modes occurs at $B < 7.5$ T. This trend is at odds with the theoretical expectation of a decreasing SW velocity with decreasing $B$, as discussed in the main text. We suspect that a weakened $\nu_B = 2$ at lower field reduces the spin polarization of the bulk and the blueshift of the modes share the same origin as the $\nu_B$ dependence shown in (a) and (b). A microscopic understanding of the non-equilibrium emission and detection process including dissipation channels can help understand these observations.

# APPENDIX H: TRANSMISSION OF GAPPED SPIN WAVE THROUGH A FM/CAF JUNCTION

The use of a contact adjacent to the CAF phase is essential to the excitation of gapless SWs. When both source and drain contacts are distant, as illustrated in Fig. 17(a), we observe gapped SW signals (Fig. 17(b)). In this setup, the SW is launched/reflected in the bulk of a FM and transmitted through a FM/CAF junction. As Fig. 17(b) shows, the envelope and symmetry of $dV_{NL}/dV$ is characteristic of a ferromagnet, where the signal onsets at a finite dc bias $V_T$. Figures 17(c) and 17(d) plot the evolution of $dV_{NL}/dV$ with increasing $B$-field and the $B$-dependence of $V_T$ respectively. $V_T$ increases linearly with $B$ with a slope of 0.09 mV/T, which is consistent with the contribution of $E_z$ ($g = 2$ gives 0.11 mV/T), but saturates to a value of ~ 1.7 mV at field below 10 T. We again observe pronounced and reproducible oscillations in $dV_{NL}/dV$. Figure 17(e) compares the oscillations in Fig. 17(b) with that of the gapless SW signal in Fig. 4(b). The remarkable correspondence of the two profiles allows us to identify the resonant modes of the CAF cavity occurring at quantized *longitudinal* momentum $k_x = n\pi/w$. Figure 17(f) plots the dc bias of the modes $V_n$ vs $n$. $V_n$ extrapolates to a finite gap at $n = 0$. Its rapid change with $n$ translates to a strong dependence of $\omega$ on $k_x$.

Several aspects of the data can be understood by recognizing the effect of energy $\omega$ and transverse momentum $k_y$ conservation on the SW transmission at the $\nu = 2 / \nu = 0$ junction (*41*). As illustrated in the inset of Fig. 17(f), the dispersion in the $\nu = 2$ region follows $\omega'(k_{x'}, k_y) = E_z + a(k_{x'}^2 + k_y^2)$ while in the CAF region $\omega(k_x, k_y) = v_{af}k = v_{af}\sqrt{k_x^2 + k_y^2}$ with $k_x$ given by $k_x = n\pi/w$. Energy conservation dictates a large $k_y$ when $k_x$ is small, which in turns increases the Zeeman gap by $ak_y^2$. This additional term could increase $V_T$ beyond the Zeeman term. The strong dependence of $\omega$ on $k_x$ our data show likely originates from the involvement of a large $k_y$, which means the SW is incident at the interface with a small angle $\theta$ as illustrated in Fig. 17(a). This scenario is supported by the dimensions of the device and the measurement setup. A quantitative understanding of transmission at heterojunctions will enable the design of magnonic devices (*42*) and further unleash the power of SW excitation as a useful tool to probe fundamental phenomena of magnetic systems.

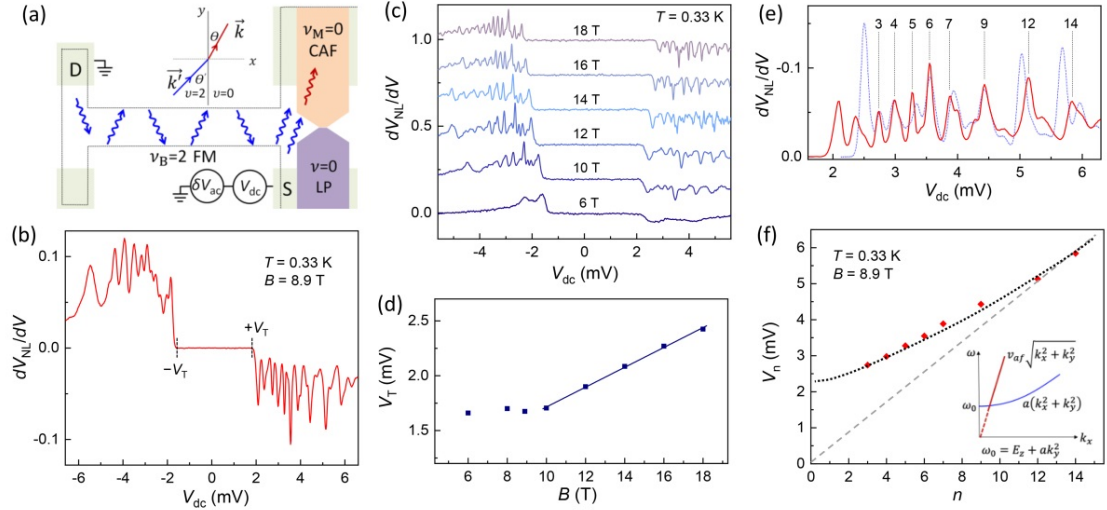

Fig. 17. SW transmission through a FM/CAF junction. (a) The measurement setup. At the $\nu = 2 / \nu = 0$ interface, $\omega = \omega'$ and $k_y = k_{y'}$. (b) $dV_{NL}/dV$ obtained in the setup shown in (a). The dashed lines mark the bias threshold $V_T$. (c) plots $dV_{NL}/dV$ obtained at selected magnetic fields as labeled. (d) plots the $B$-dependence of the threshold $V_T$ averaged for $\pm V_{dc}$ and a linear fit to the data with a slope of 0.09 mV/T. (e) overlays the data in (b) ($V_{dc} > 0$, red solid line) with the corresponding trace in Fig. 4(b) (blue dotted line, expanded in $V_{dc}$

by × 4.55 and shifted horizontally). The remarkable overlap indicates that both arise from the FP resonance of the CAF region. (f) plots $V_n$ vs $n$ extracted from (e). The dotted line is a guide to the eye. From device 606. The gray dashed line has a slope of 0.42 mV/mode, which is 7 times of the low-$T$ slope in Fig. 4(e). This indicates $k_y \gg k_x$ and a shallow SW incidence as illustrated in (a). The inset of (f) illustrates the $\omega - k_x$ relations in the FM (blue line) and the CAF (red line) regions. The CAF phase also supports a gapped dispersion with $\omega_0 = 2E_z$ (*12*) but this branch is not activated due to energy conservation.

## Acknowledgements

H.F., K.H. and J.Z. are supported by the Kaufman New Initiative research Grant No. KA2018-98553 of the Pittsburgh Foundation and the National Science Foundation Grant No. NSF-DMR-1904986. H.F. acknowledges the support of the Penn State Eberly Postdoctoral Fellowship. K.W. and T.T. acknowledge support from the Elemental Strategy Initiative conducted by the MEXT, Japan, Grant Number JPMXP0112101001, JSPS KAKENHI Grant Number JP20H00354 and the CREST (JPMJCR15F3), JST. Part of this work was performed at the NHMFL, which is supported by the NSF through NSF-DMR-1644779 and the State of Florida. We are grateful for helpful discussions with So Takei, Luqiao Liu, Chunli Huang, Nemin Wei, Allan H. McDonald, Jainendra Jain, Bertrand Halperin and Falko Pientka.

## Competing interests

The authors declare no competing interests.